\documentstyle[twoside]{memsait}

\begin{opening}
\title{HOW WELL DO STANDARD SOLAR MODELS DESCRIBE THE RESULTS OF SOLAR
NEUTRINO EXPERIMENTS?\thanks{Invited talk at the symposium on {\it The
Inconstant Sun}, Naples, Italy, March 18, 1996.  To be published in
{\it Memorie della Societ\'a.}, eds. G. Cauzzi and C. Marmolino.}}
\author{JOHN BAHCALL}
\institute{Institute for Advanced Study, Princeton, NJ 08540}
\date{}
\end{opening}

\begin{document}
\input psfig
\baselineskip=24pt

\oddpagefooter{}{}{} 
\evenpagefooter{}{}{} 

\bigskip

\begin{abstract}
The neutrino fluxes calculated from 
the 14 standard solar models published recently in refereed
journals are inconsistent with the results of the 4
pioneering solar neutrino experiments if nothing happens to the
neutrinos after they are created in the solar interior. 
The calculated fluxes and the experimental results are in good
agreement if neutrino oscillations occur.
\end{abstract}

\section{Introduction}
\label{intro}

Solar neutrino research has achieved its primary goal, the 
detection of solar neutrinos, and is now entering a new phase in which
large electronic detectors will yield vast amounts of diagnostic data.
The new experiments (Arpesella et al. 1992, Takita 1993, McDonald
1994)  will focus on testing the prediction of standard
electroweak theory (Glashow 1961, Weinberg 1967, Salam 1968)
that essentially nothing happens to electron type
neutrinos after they are created by nuclear fusion reactions in the
interior of the sun.
The purpose of this talk is, on the eve of the new experiments, 
 to assess the results of three decades of confrontation between solar
models and solar neutrino experiments  and to indicate some of the
 challenges that 
lie ahead.

The four pioneering experiments--chlorine (Davis 1964, 1994), which 
uses $\rm {C_2Cl_4}$ as a detector, 
Kamiokande (Suzuki 1995), a water Cerenkov experiment, 
GALLEX (Anselmann et al. 1995), and SAGE (Abdurashitov et
al. 1994), gallium radiochemical experiments--have 
all observed neutrino fluxes with intensities that are within a
factors of a few of 
those predicted by standard solar models. 
Three of the experiments (chlorine , GALLEX, and SAGE) are
radiochemical and each radiochemical experiment  measures
 one number, the total rate at which
neutrinos above a fixed energy threshold (which depends upon the
detector)  are captured.  
The sole electronic detector among the initial experiments,
Kamiokande, has shown 
that the neutrinos come from the sun,
by measuring the recoil directions of the  electrons scattered by
solar neutrinos, and has also demonstrated 
that the neutrino energies
are in the range expected on the basis of the standard solar model.

Despite continual refinement of solar model calculations of
neutrino fluxes over the past 35 years (see, e.g., the collection of
reprint articles in Bahcall, Davis, Parker, Smirnov, and Ulrich 1995),
the discrepancies between 
observations and calculations have gotten worse with time.  All four
of the pioneering solar neutrino experiments yield event rates that
are significantly less than predicted by standard solar models.
Moreover, there are well known inconsistencies between the different
experiments if the observations are interpreted assuming that nothing
happens to the neutrinos after they are created.

In this talk, I will first summarize the results of all the 
recently published
standard solar model calculations and compare them with the results of
the four solar neutrino experiments.  For purposes of the summary, 
I will assume that, as implied
by standard electroweak theory, nothing happens to the neutrinos after
they are created.  Then I will recall  the results of many
authors which show 
 that the results of the solar neutrino experiments can be
explained well if neutrinos oscillate between different eigenstates,
i.e., between different types of neutrinos.  Finally, I will discuss
the implications for astronomy of the neutrino experiments.

\section{Observation versus Calculation}
\label{comparison}

\begin{figure}
\centerline{\psfig{figure=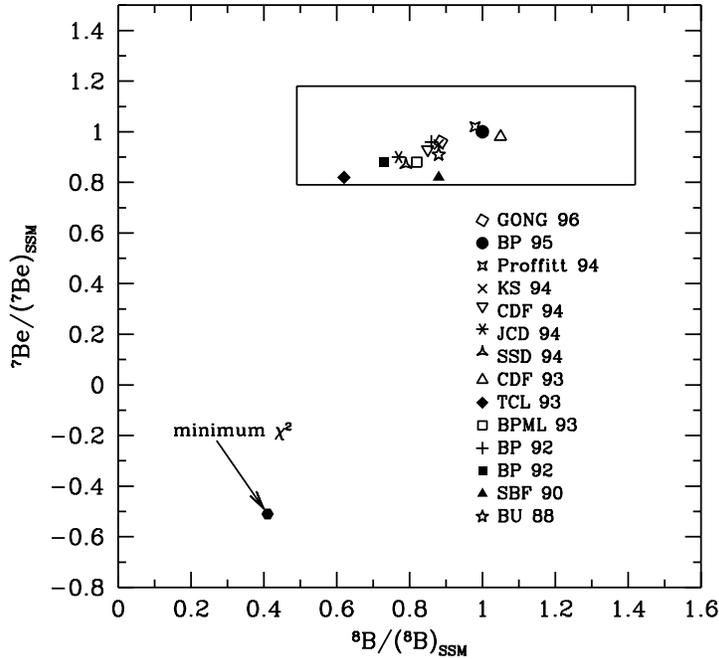,width=4in}}
\vglue-.2in
\caption[]{The calculated ${\rm ^7Be}$ and ${\rm
^8B}$ solar neutrino fluxes for all 14 of the standard solar models.
All of the fluxes have been normalized by dividing by the Bahcall and
Pinsonneault (1995) standard solar model (SSM) values. 
The abbreviations of the various solar models are GONG
(Christensen-Dalsgaard et al. 1996), BP 95 (Bahcall and 
Pinsonneault 1995),  KS 94 (Kovetz and Shaviv 
1994), CDF 94 (Castellani, Degl'Innocenti, Fiorentini, Lissia, and
Ricci 1994), JCD 94
(Christensen-Dalsgaard 1994), SSD 94 (Shi, Schramm, and Dearborn
1994), CDF 93  (Castellani, Degl'Innocenti, and Fiorentini 1993), TCL
93 (Turck-Chi\`eze and Lopes 1993), BPML 93 (Berthomieu, Provost,
Morel, and Lebreton 1993), BP 92 (Bahcall and Pinsonneault 1992), SBF 90 (Sackman,
Boothroyd, and Fowler 1990), and BU 88 (Bahcall and Ulrich 1988).}
\label{fluxes}
\end{figure}

Figure~\ref{fluxes} displays the calculated ${\rm ^7Be}$ and ${\rm
^8B}$ solar neutrino fluxes for all 14 of the standard solar models
with which I am  familiar that have been published in refereed
science journals since 1988 (and before the cutoff date for this
review: June 1, 1996).  The first systematic discussion of the
relation between helioseismology and solar neutrino research was
published in 1988 (Bahcall and Ulrich 1988). I normalize the fluxes by dividing each
published value by the flux from the most recent 
Bahcall and Pinsonneault (1995)
standard
solar model (hereafter BP95) which makes use of improved 
input parameters and
includes heavy element and helium diffusion.  The abscissa is the
normalized ${\rm ^8B}$ flux and the numerator
 is the normalized ${\rm ^7Be}$
neutrino flux.  The box shows the estimated 
3$\sigma$ uncertainties in the
predictions of the standard solar model (BP95).
The abbreviations that indicate references to individual models are
identified in the caption of Figure~\ref{fluxes}. 

All of the solar model results from different groups 
fall within the estimated 3$\sigma$
uncertainties in the model predictions.  
This agreement between the results of 14 groups 
demonstrates the robustness of the predictions since the
 calculations use different computer codes and 
involve a variety of choices for the nuclear
parameters, the equation of state, the stellar radiative opacity, 
the initial heavy element abundances, and the physical processes
that are included.
In fact, all published standard  solar
models give the same results for solar neutrino fluxes to an accuracy
of better than 10\% if the same input parameters and physical
processes are included (Bahcall and Pinsonneault 1992, 1995)

The largest contribution to  the dispersion in values in
Figure~\ref{fluxes} is
caused by  the inclusion, or non-inclusion, of element diffusion in the 
stellar evolution codes.
The Proffitt (1994), the
Bahcall and Pinsonneault (1995), and the Christensen-Dalsgaard et al.
(1996)  models all  include helium and heavy element
diffusion.  The 
predicted fluxes in these three models agree to within $\pm 10\%$,
although the  models
are calculated  using different mathematical descriptions of
diffusion (and somewhat different input parameters),
The calculated value  that is furtherest from the center of the box 
is by Turck-Chi\`eze and Lopes (1993), which does not include
either helium or heavy element diffusion.  However, the Turck-Chi\`eze
and Lopes best estimate is still well within the 3$\sigma$ box.

Helioseismology has recently sharpened
the disagreement between observations and the predictions of 
solar models 
with standard (non-oscillating) neutrinos.
By including element diffusion, 
four solar models  near the center of the box in
Figure~\ref{fluxes} (models of Bahcall and Pinsonneault 1992, 
Proffit 1994, 
BP95, and Christensen-Daalsgard et al. 1996)
yield values for the depth of the convective zone  and the primordial
helium abundance that are in agreement with helioseismological
measurements. (The model of Richard et al. 1996 yields results in good
agreement with the four solar models just mentioned that include
element diffusion, but was not yet published in Astron. and Astrophys.
by the cutoff date, June 1, 1996.)

Solar models that do not include diffusion 
are not consistent with the helioseismological evidence (see discussion in
Christensen-Dalsgaard, Proffitt, and Thompson 1993, Guzik and Cox
1993, BP95, and
Christensen-Daalsgard et al. 1996).  The results of the major
new helioseismological initiatives,  GONG and SOHO, will provide important
additional constraints on the solar models.

In my view, only solar models that include element diffusion should,
in the future,  be
called ``standard solar models''. 
These ``standard models'' all lie close to the center of the
rectangular error box in Figure~\ref{fluxes}.
 The physics of
diffusion is simple and there is an exportable subroutine available
for calculating diffusion in stars (see
http://www.sns.ias.edu/$^\sim$jnb).  Observation requires, and
computing technology 
easily permits, the inclusion of diffusion in any standard
stellar evolution code. 

How do the 
observations from the four pioneering solar neutrino
experiments agree with the solar model calculation?  
Plamen Krastev and I (see Bahcall and Krastev 1996 for a
description of the techniques) have recently compared the
predicted standard model fluxes, with their estimated uncertainties,
and the observed rates in the chlorine, Kamiokande, GALLEX, and SAGE
experiments.  The theoretical solar model and experimental
uncertainties, as well as the uncertainties in the neutrino cross
sections, have been
combined quadratically. Using the predicted fluxes from the BP95
model, the $\chi^2$ for the fit to the four experiments is

\begin{equation}
\chi^2_{\rm SSM} \hbox{(all 4 experiments)} = 56\ .
\label{chitofit}
\end{equation}
The theoretical uncertainties (from the solar model and the neutrino
cross section calculations) and the experimental errors (statistical
and systematic) have been combined quadratically in obtaining equation
(\ref{chitofit}).

Suppose we now ignore what we  have learned  from solar models and allow
the important 
${\rm ^7Be}$ and ${\rm ^8B}$ fluxes to take on any non-negative
values.  What is the minimum value of $\chi^2$ for the 4 experiments,
when the only constraint on the fluxes is the requirement that the
luminosity of the sun be supplied by nuclear fusion reactions among
light elements?  We include the nuclear physics inequalities between
neutrino fluxes (see section 4 of Bahcall and Krastev 1996) that are
associated with the luminosity constraint and maintain the standard
value for the almost
model-independent  ratio of $pep$ to $pp$ neutrinos.

The best fit for arbitrary $^7$Be and $^8$B neutrino fluxes 
is obtained for ${\rm ^7Be/(^7Be)}_{\rm SSM} = 0$ and
${\rm ^8B/(^8B)}_{\rm SSM} = 0.40$, where

\begin{equation}
\chi^2_{\rm minimum} \hbox{(all 4
experiments; ~arbitrary $^7$Be, $^8$B)} = 14.4\ . 
\label{chiall}
\end{equation}
The CNO neutrinos were assumed equal to their standard model values in
the calculations that led to Eq.~\ref{chiall}.  The fit can be further
improved if we set the CNO neutrino fluxes equal to zero. 
Then, the same search for arbitrary $^7$Be and
$^8$B neutrino fluxes leads to 

\begin{equation}
\chi^2_{\rm minimum} \hbox{(all 4
experiments; ~arbitrary $^7$Be, $^8$B; CNO = 0)} = 5.9\ .
\label{chinoCNO}
\end{equation}

If we drop the physical requirement that the fluxes be positive
definite, the minimum $\chi^2$ occurs (cf. 
Figure~\ref{fluxes}) for a negative value of the ${\rm ^7Be}$
flux; this unphysical 
result is a reflection of what has become  known in the physics
literature as `` the  missing ${\rm
^7Be}$ solar neutrinos.''.  The reason that the $^7$Be neutrinos appear
to be missing (or have a negative flux) is that the two gallium
experiments, GALLEX and SAGE, have an average event rate of $74 \pm 8$
SNU,
which is fully accounted for in the standard model 
by the fundamental $p-p$ and $pep$ 
neutrinos (best estimate $73 \pm 1$ SNU). 
In addition, the $^8$B neutrinos
that are observed in the Kamiokande experiment will produce about $7$
SNU in the gallium experiments, unless new particle physics affects
the neutrinos.

To me, these results suggest strongly that the assumption on
which they are based---nothing happens to the neutrinos after they are
created in the interior of the sun---is incorrect.  A less plausible 
 alternative
(in my view)  is that some of the
experiments are wrong; this must be checked by further experiments.

\section{Are Neutrino Oscillations the Answer?}
\label{oscillations}

In the simplest version of the 
standard model of electroweak interactions (Glashow 1961,
Weinberg 1967, Salam  1968), electron-type
neutrinos that are created in the center of the sun by nuclear fusion
reactions remain electron-type neutrinos as they pass through the
solar material and propagate to detectors on earth. The three 
radiochemical
experiments (chlorine, GALLEX, and SAGE) are sensitive only to
electron-type neutrinos, whereas Kamiokande has reduced
sensitivity also for muon or tau neutrinos.

Particle physicists have proposed a number of possible solutions to
the problem posed by the discrepancy between solar neutrino 
observations and 
the combined standard predictions of solar models and 
electroweak theory.  The most popular of these
solutions involve  neutrino oscillations in vacuum (Pontecorvo 1968)
and matter enhanced resonant neutrino oscillations, the so-called MSW
effect (Wolfenstein 1978, Mikheyev and Smirnov 1985). 

The comparison between theory and observations 
is improved significantly if neutrino oscillations occur.  
I give here the results of calculations 
for the particle physics solutions that are
most frequently discussed in the physics literature.
The
minimum $\chi^2$ obtained with two degrees of freedom (mixing angle,
and difference of squared masses) is (Bahcall and Krastev 1996), 
for the most-popular small mixing angle 
Mikheyev-Smirnov-Wolfenstein (MSW) solution,

\begin{equation}
\chi^2_{\min} = 0.31\ \ \ ,\ \ \ {\rm SMA}\ .
\label{chioscSMA}
\end{equation}
For the large mixing angle (MSW) solution,
\begin{equation}
\chi^2_{\min, } = 2.5\ \ \ ,\ \ \ {\rm LMA}\ .
\label{chioscLMA}
\end{equation}
For vacuum neutrino oscillations,
\begin{equation}
\chi^2_{\min}= 2.5\ \ \ ,\ \ \ {\rm vacuum~oscillations}\ .
\label{chioscvac}
\end{equation}

Neutrino oscillations provide a significant improvement in
the minimum $\chi^2$ for the four operating solar neutrino
experiments.

\section{Discussion}
\label{closing}

The combined predictions of the standard solar model and the standard
electroweak theory disagree with the results of the four pioneering
solar neutrino experiments.  
Comparing the combined predictions to the existing data, we obtain
  values for
$\chi^2_{standard}$ of  $\sim 56$.
The fits are
much improved if neutrino oscillations, which are described by  two free
parameters, are included in the calculations.  With neutrino
oscillations, the characteristic value for $\chi^2_{\min, ~osc.} \sim
1$.  New experiments (Arpesella et al. 1992, Takita 1993, McDonald
1994) involving large electronic detectors of
individual neutrino events will decide in the next few years if
neutrino oscillations are indeed important in interpreting solar
neutrino experiments.

For astrophysics, the most important quantities that can be deduced from
neutrino oscillation experiments are the neutrino mass differences
(Only the squares of mass differences appear in the oscillation
equations, since the propagation phases are determined by the squares
of masses.).  
For the currently most popular oscillation scenario, the MSW effect
(Mikheyev and Smirnov 1985; Wolfenstein 1978) [which involves resonant
flavor conversion in matter],
the values of the mass differences reported in the literature 
are obtained by
solving the differential equations for neutrino
 propagation  in
matter.

There is  a simple analytic argument which
allows one to estimate the neutrino masses that result from numerical
solutions of the MSW propagation equations and to understand why the
neutrino masses are given robustly by MSW theory.  Let $n_e$,
$\theta_V$,
$\Delta m^2$, and $E_\nu$ be, respectively, the electron number
density, the mixing angle in vacuum between two types of neutrino
states (e.g., electron type and muon type), the difference of the
squared masses of the two different neutrino types, and the neutrino
energy. Then one can show analytically (Mikheyev and Smirnov 1985) 
that there is a resonance in
the neutrino propagation only if somewhere in the sun the electron
density at resonance satisfies the  following numerical equation 
(Eq. 9.53 of Bahcall 1989):

\begin{equation}
\frac{n_e(\hbox{resonance)}}{n_e\hbox{(center of sun)}} = 0.7\cos
2\theta_V \left[\frac{\Delta m^2}{10^{-5} {\rm eV^2} }\right]
\left[ \frac{\hbox{1 MeV}}{E_\nu}\right].
\label{resonance}
\end{equation}
Obviously, there is no solution to Eq.~(\ref{resonance}) if the
required value for $n_e(\hbox{resonance)}$ exceeds the highest value
of the electron density, which occurs at the center of the sun.

As remarked in Section~\ref{comparison}, the two gallium experiments
suggest that the $p-p$ neutrinos (with energies less than $0.4$ MeV)
are not affected by resonance oscillations while the $0.86$ MeV $^7$Be
neutrinos are affected by the resonance.  
Requiring that ${n_e(\hbox{resonance)}}/{n_e\hbox{(center of
sun)}}$ be greater than unity for $E_\nu = 0.4$ MeV and less than
unity for $E_\nu = 0.9$ MeV, yields

\begin{equation}
\Delta m^2 \sim 10^{-5} \ \ {eV^2} .
\label{MSWmass}
\end{equation}
It is plausible to suppose that 
Eq.~(\ref{MSWmass}) gives approximately the mass of the muon
neutrino (i.e., 
$m(\nu_\mu) \sim 0.003$ eV), which is expected to be heavier than 
the electron neutrino.  
Many particle physics
models suggest that the mass of the tau neutrino is larger than
the mass of the muon neutrino by a
factor whose order of magnitude is 
 the ratio of the square of the mass of the top quark ($176$ GeV)
to the square of the mass of the charmed quark ($1.6$ GeV).  
One might anticipate, therefore, a mass for the tau neutrino 
that is within a factor of ten of 
 $10^4 \times 0.003$ eV, or 

\begin{equation}
m(\nu_\tau) ~\sim~ 10^{1.5} \ \ eV .
\label{taunumass}
\end{equation}
This mass for the tau neutrino 
would be cosmologically important, potentially containing enough dark
matter to close the universe.

Finally, we may ask:  What have solar neutrino experiments taught us
about astronomy? The operating 
experiments have achieved the primary goal of
solar neutrino astronomy by showing empirically that the sun shines
via nuclear fusion reactions.  Moreover, the observed and the standard
predicted
neutrino interaction rates agree within  factors of a few, providing
(see below)
semi-quantitative confirmation of the calculation of
temperature-sensitive nuclear fusion rates
in the solar interior.

The important $^8$B neutrino flux depends 
upon the central temperature of the sun as approximately $T^{24}$ 
(Bahcall and Ulmer 1996).  The maximum range allowed by neutrino
oscillation scenarios and the results of the four operating solar
neutrino experiments is (Bahcall and Krastev 1996):

\begin{equation}
0.37 \leq \phi {\rm (^8B)}/\phi {\rm (^8B)}_{\rm SSM} \leq 2.84\ .
\label{maxrange}
\end{equation}
Thus the Kamiokande experiment constrains the total ${\rm ^8B}$
neutrino flux to be within a factor of three of the value predicted by
standard solar models (if neutrino oscillations, vacuum or resonant
matter oscillations are occurring).

The possibility that neutrino oscillations are occuring
complicates greatly the interpretation of solar neutrino 
data.  Until new experiments are performed, one 
cannot even rule out empirically an {\it ad hoc} scenario (Bahcall,
Fukugita, and Krastev 1996), not predicted by any detailed solar
model, in which the sun shines by CNO rather than $p$-$p$ fusion reactions.

The  SNO (McDonald 1994) heavy water 
experiment
will measure for $^8$B solar neutrinos 
both the total flux and the flux of electron type neutrinos.
The Superkamiokande ultrapure water 
experiment (Takita 1993), which began operating
April 1, 1996, is primarily sensitive to electron type neutrinos but
has some sensitivity to other neutrino types also.
The results of these experiments 
 will determine the absolute
value of the $^8$B neutrino production  rate in the sun, 
which was the originally-stated purpose of the chlorine experiment 
(Bahcall 1964, Davis 1964) before the
complications due to possible new neutrino physics were recognized. 
The results from these new experiments will constitute a critical,
quantitative test, independent of uncertainties about new particle
physics,  of solar model calculations of nuclear fusion rates
in the center of the sun.

\acknowledgements
I am grateful to Plamen Krastev for valuable conversations and for
performing the $\chi^2$ calculations and to E. Lisi for helpful
comments.  The research of John Bahcall is supported in part by NSF
grant number PHY95-13835.

\end{document}